# Extracting Biomedical Factual Knowledge Using Pretrained Language Model and Electronic Health Record Context

Zonghai Yao, MSc[1], Yi Cao, BS[1], Zhichao Yang, MSc[1],
Vijeta Deshpande, MSc[2], Hong Yu, PhD[1,2,3,4]

[1] College of Information and Computer Science, University of Massachusetts Amherst, Amherst, MA, USA; [2] Department of Computer Science, University of Massachusetts Lowell, Lowell, MA, USA; [3] Department of Medicine, University of Massachusetts Medical School, Worcester, MA, USA; [4] Center for Healthcare Organization and Implementation Research, Bedford Veterans Affairs Medical Center, Bedford, MA, USA

**Abstract**

*Language Models (LMs) have performed well on biomedical natural language processing applications. In this study, we conducted some experiments to use prompt methods to extract knowledge from LMs as new knowledge Bases (LMs as KBs). However, prompting can only be used as a low bound for knowledge extraction, and perform particularly poorly on biomedical domain KBs. In order to make LMs as KBs more in line with the actual application scenarios of the biomedical domain, we specifically add EHR notes as context to the prompt to improve the low bound in the biomedical domain. We design and validate a series of experiments for our Dynamic-Context-BioLAMA task. Our experiments show that the knowledge possessed by those language models can distinguish the correct knowledge from the noise knowledge in the EHR notes, and such distinguishing ability can also be used as a new metric to evaluate the amount of knowledge possessed by the model.*

## 1. Introduction

Pretrained language models (LMs), such as ELMo[1], BERT[2], and RoBERTa[3], have achieved incredible success on many downstream tasks. The researchers found that these LMs were able to predict either the next word in a sequence or masked words anywhere in a given sequence (e.g. "Tinea corporis (disorder) has symptoms such as [Mask]."). The parameters of these models appear to store a large amount of Linguistic knowledge useful for downstream NLP tasks[4]. These inspired researchers to measure how much factual information LMs can obtain from their pretraining time. LAMA[5] formally defines this task. For each triple (subject, relation, object) in the knowledge base, LAMA will design the human-written templates to express each relation. They show that LMs like BERT can predict objects given cloze-style prompts. For example, "The team with the most World Cup wins is [MASK]." In the biomedical domain, there has been recent work focusing on factual knowledge probing for the biomedical domain and releasing the Biomedical LAnguage Model Analysis (BioLAMA) probe[10].

However, while the paradigm of prompt has been used to achieve some interesting results about knowledge expressed by LMs, they often rely on prompts that are created manually based on the experimenter's intuition[6, 7], or generated by some algorithm[8, 9]. Regardless of the existing methods, no one can guarantee that the prompts used (e.g. "Barack Obama was born in") are optimal, since the LMs may have learned the target knowledge from vastly different contexts during training[5] (e.g. "Barack Obama was born in Honolulu, Hawaii."). Since the hint is not a valid query for that fact, there is a good chance that a fact that the LMs does know cannot be retrieved. In fact the LMs may be richer than these initial results suggested[5]. In this case, LAMA presented the existing results only as a lower bound on the range of knowledge contained in the LMs and subsequent work[6, 7, 8, 9] has attempted to tighten this bound by finding better prompts. This problem is more severe in N-to-one relation or N-to-M relations[10] (e.g., diseases to symptoms). Because if it is just a one-to-one relation, this relation is more likely to be mentioned in many different contexts during the pretraining process (meaning that it is easy to be retrieved by different prompts); but if we think of N-to-one or N-to-M as many one-to-one relations, then each relation is mentioned less, which means that each one-to-one can only be used by a specific relation prompt retrieved. This makes it more difficult to guarantee that the corresponding knowledge can be obtained with the prompt. Since N-to-1 and N-to-M relations are very common in biomedical KBs, we need new methods to improve this low bound.

Besides, the ultimate goal of this work is to make LMs more reliable KBs[5]. In the actual usage scenario of KB, how to give the correct knowledge based on some specific content is more meaningful and more accurate to meet the requirements. For example, in EHR notes, each disease has many different symptoms. Few physicians can list all the symptoms of all diseases, yet they are often able to make a diagnosis based on the descriptions of various symptoms in the EHR notes. Therefore, our requirement for LMs should not simply enumerate all the symptoms of a certain disease, but whether it can find useful knowledge based on specific symptoms in the notes.

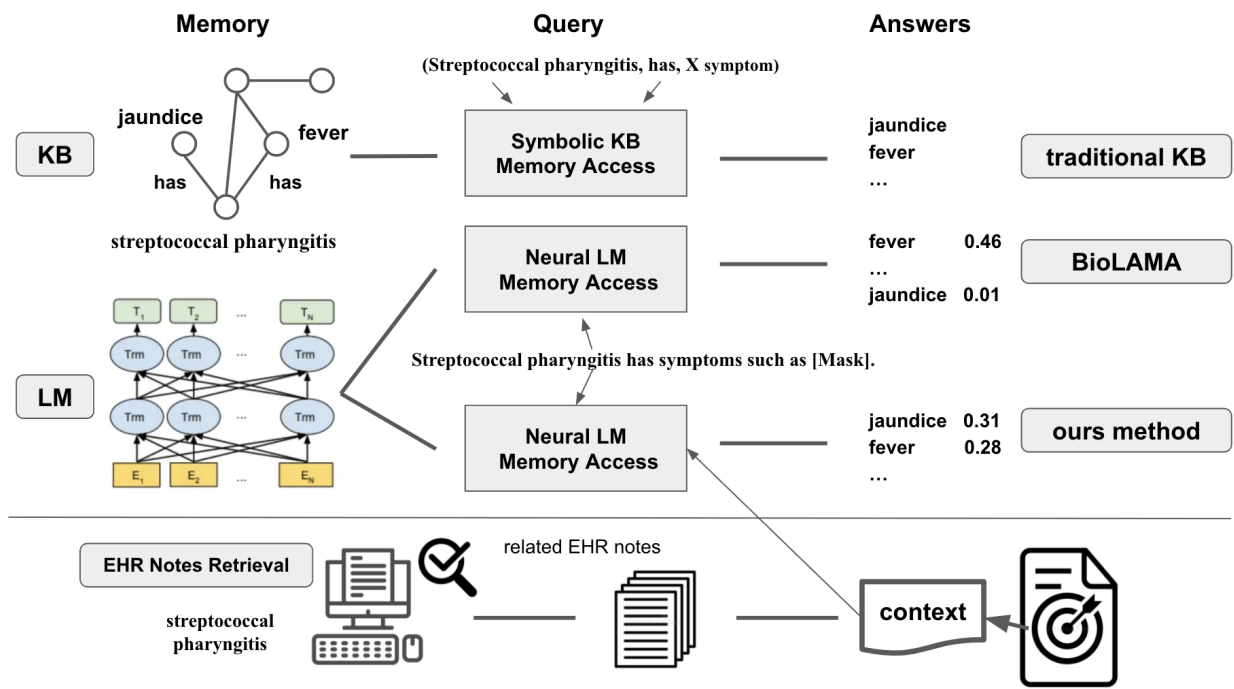

**Figure 1.** Querying knowledge bases (KB) and language models (LM) for factual knowledge. We add the relevant EHR note as a context to the prompt to raise the lower bound of this prompt paradigm in extracting LM knowledge

The above reasons inspired us to add real EHR note data to the prompt as an important and indispensable context to extract and verify knowledge in LMs. In this paper we focus on the relationship between diseases and symptoms to design experiments. The reason we choose EHR notes as the context is because there are a large number of correct and incorrect relations in these notes. For example, a patient's Assessment has the diagnosis, which is typically with its disease name, and the Subjective part in EHR will mention a large number of symptoms related and unrelated to this disease. Due to local attention[11], the LM has a higher probability of mentioning the symptoms that appear in the context when doing "filling in the blank" task. Under this setting, we "soft-constrict" the candidate symptoms to the ones mentioned in the EHR note context, and we design a set of rigorous and reasonable experiments to evaluate whether the LM can give the correct symptoms higher ranking based on its existing knowledge. Our results show that after adding the EHR context, the scores for the prompt paradigm are greatly improved. Further results demonstrate that the knowledge possessed by the model can distinguish the correct knowledge (correct symptoms) from the noise knowledge (incorrect symptoms) in the EHR, and such distinguishing ability can also be used as a metric of the amount of knowledge possessed by the model.

## 2.Related Work

**Knowledge Bases (KBs)** Knowledge bases are one solution to access annotated ground truth relational data by enabling queries (Tinea corporis (disorder), has, X symptom). However, in practice, we often need to extract structured relational data from a variety of sources to populate these knowledge bases. This requires complex NLP pipelines including named entity recognition[12], coreference resolution[13], entity linking[14, 15], and relationship recognition[16, 17], which leads to errors that can easily propagate and accumulate throughout the pipeline. Instead, we can try to query relational data by asking the neural language model to populate mask tokens in sequences like "Tinea corporis (disorder) has symptoms such as [Mask]", as shown in Figure 1. In this context, language models have various attractive properties: they do not require schema engineering, do not require human annotation, and support an open set of queries. In this paper, we hope to follow the previous work to explore the next generation of knowledge base solutions, such as: Can we get the desired answer directly in the form of human natural language like asking an experienced expert? This requires us to see enough convincing evidence through tasks such as LAMA to prove that LMs can carry such an amount of knowledge.

**LMs as KBs** The factual probing task was introduced by the LAMA benchmark, which aims to measure the amount of factual information encoded in LMs. In LAMA, facts are defined as triples (s, r, o), where s is a subject (e.g., Tinea corporis (disorder)), r is a relation from a fixed set of relations R (e.g., disease-symptom), and o is an object (e.g., Itch). In the LAMA setting, each relationship is associated with a human-written prompt containing a single [MASK] token. Petroni et al, (2019)[9] state that their benchmarks only provide a lower bound estimate of the amount of factual information stored in the LM, as their hand-written prompts may not be optimal for eliciting facts. Therefore, subsequent work focuses on tightening this bound by using additional training data to find better prompting methods[6, 7, 8, 9]. However, previous work mostly tried to improve the quality of the prompt to raise the lower bound of this prompt paradigm in extracting LMs' knowledge. Our goal is also to raise the lower bound. However, different from the previous methods, we hope to achieve the goal by combining the real application scenarios for Biomedical KBs with the local attention mechanism of LMs. Specifically, we concat the EHR note context with the prompt, and when the LMs do "filling in the blank" task according to the input, its output will be "soft-constrained" to candidates that appear in context (such as all symptoms appearing in the EHR note). Such local attention makes the LM pay more attention to the content in these contexts. We will evaluate whether such soft constraint help improve biomedical KBs from LMs.

**BioLAMA** In the biomedical domain, Sung et al, (2021)[10] created and released the BioLAMA probe following the LAMA setting. BioLAMA consists of 49K biomedical factual triples whose relations have been manually curated from three different knowledge sources: the Comparative Toxicogenomics Database (CTD), the Unified Medical Language System (UMLS), and Wikidata. Therefore, we evaluate our models on the BioLAMA benchmark.

## 3.Methods

**Context (EHR Notes) Retrieval** We hope to use the structural information of the EHR note to find a reasonable context for triples of (subject, relation, object) without expert annotation, so in this paper we take the diseases-symptoms relation as an example to introduce our heuristic information retrieval method. The EHR notes we use is UMMS SOAP DATABASE because it has a clear SOAP structure, i.e. (Subjective, Objective, Assessment, Plan). For a certain triples (disease_A, has, symptom_B), we first retrieve all EHR notes in DATABASE to get **D1**. The retrieval condition is that the Assessment section has and only has disease_A. The reason for "has and only has" is that we hope that this disease is the diagnosis in this note, rather than being mentioned by other reasons, so as to ensure that the context of the context can infer the corresponding (disease, has, symptom) relation. We continue to retrieve D1 to get **D2**, and the retrieval condition is that the Subject section has this symptom_B.

**Dynamic-Context-BioLAMA** In order to more deeply and comprehensively understand and analyze whether the model can use context to help itself better "say" knowledge in the BioLAMA task, we added the EHR note context in D2 to the prompt data in BioLAMA. We observe how the model responds to different symptoms in the context by strictly controlling the window size (WS) of the context. Here we define the WS of the context. For a triple (disease_A, has, symptom_B) and an EHR note in D2, we use the NER tool to find all the symptoms in the Subjective section. Taking the position of symptom_B as the center, we divide the context into different segments according to these symptoms. Taking symptom_B as the center, we add segments to both sides in turn. Each time a segment is added, it will bring us a new symptom and some related text.

**EHR note**
**Subjective**: [segment11] There is intermittent cough in the morning. There is no more hemoptysis [segment10] in the past week. There is no shortness of breath [segment9] , wheeze [segment8] , chest pain [segment7] , chest tightness [segment6] , hemoptysis [segment4] , orthopnea or paroxysmal nocturnal dyspnea. There is no heartburn [segment2] , sour taste in the mouth or regurgitation, warm sick feeling [segment1] in the chest or difficulty swallowing. There is no sense of smell or taste. There is discolored nasal discharge on the right and on the left. There is [segment3] postnasal drip and throat clearing, and [segment5] nasal stuffiness on the left.
**Assessment**: The patient has residual nasal polyp on the left.

**Prompt:** nasal polyp has symptoms such as [Mask].

S1 + Prompt → LM
S2 S1 + Prompt → LM
S2 S1 S3 + Prompt → LM
S4 S2 S1 S3 + Prompt → LM
S4 S2 S1 S3 S5 + Prompt → LM
S6 S4 S2 S1 S3 S5 + Prompt → LM
S7 S6 S4 S2 S1 S3 S5 + Prompt → LM
S8 S7 S6 S4 S2 S1 S3 S5 + Prompt → LM
S9 S8 S7 S6 S4 S2 S1 S3 S5 + Prompt → LM
S10 S9 S8 S7 S6 S4 S2 S1 S3 S5 + Prompt → LM
S11 S10 S9 S8 S7 S6 S4 S2 S1 S3 S5 + Prompt → LM

**Figure 2.** Dynamic-Context-BioLAMA. For triple (disease_A, has, symptom_B), red font represents disease_A, orange font represents symptom_B, pink font represents other symptoms that are equally correct in the context, and blue font represents equally wrong symptoms in the context. We divide the EHR context into different segments according to the symptoms, and then we add the context to the prompt segment by segment to construct such dynamic inputs to observe how the output of the model changes.

In order to dynamically observe how the model responds to contexts containing "correct" or "wrong" symptoms, before and after adding new segments, we will count the following: 1) rank change of symptom_B; 2) rank of newly added symptom; 3) The average rank change of all correct symptoms currently appearing in the context; 4) The average rank change of all wrong symptoms currently appearing in the context. Here, the "rank change" will use a positive number to indicate a rank up and a negative number to indicate a rank down.

### 4.Experimental Setup

**Data Statistics** Biomedical Informatics Ontology System (BIOS)* is a comprehensive biomedical knowledge graph including diseases, symptoms, medications, surgeries and non-surgical treatments. We use BIOS as our knowledge during experiments to evaluate each model. After preprocessing BIOS data, there are 3459 diseases, each disease has 14.018 on average. The EHR notes in this article are from the UMass Memorial Hospital, which is SOAP-structured[20] (Subjective, Objective, Assessment, and Plan) data and contains a total of 18,867 EHR notes, we name it UMass-EHRs dataset. From all SOAP EHR notes, we find 7358 notes mentioning 240 diseases in our preprocessed data. The number of symptoms for each disease mentioned by these notes accounts for 17.9% symptoms for each disease on average. The average length of the prompt is 6.743 tokens, and the length after adding the context is determined by the number of segments added (as shown in Figure 2). From adding S1 to adding S7, the average length of the input (context+prompt) are 78.428, 104.408, 132.037, 157.976 , 174.204, 202.415, and 211.3 tokens.

**Language Model** We use one general-domain LM and several biomedical LMs: BERT[2], BioBERT[21], and 5 different Bio-LMs[22] which is pretrained based on the RoBERTa-base[3]. BioBERT and BioLMs are both pre-trained over PubMed, and some Bio-LMs also used MIMIC-III data. While some Bio-LMs also use a custom vocabulary learned from PubMed, BioBERT uses the same vocabulary as BERT.

**Prompt Method** We use a fill-in-the-blank cloze statement (i.e., a "prompt") for probing and choosing to directly use manual prompts[5]. Specifically, we used manual prompts created by domain experts following BioLAMA[10]. Since the majority of entities in our dataset are made up of multiple tokens, we also implement a multi-token decoding strategy following Sung et al, (2021)[10]. Among their decoding methods, we use the confidence-based method which greedily decodes output tokens sorted by the maximum logit in each token position. Note that we do not restrict our output spaces by any pre-defined sets of biomedical entities since we are more interested in how accurately the LMs contain biomedical knowledge in an unconstrained setting.

---
*https://bios.idea.edu.cn/

**Evaluation Metric** Following the previous work of BioLAMA, we use top-k accuracy (Acc@k), which is 1 if any of the top k object entities are included in the annotated object list, and is 0 otherwise. We use both Acc@1 and Acc@5 since most biomedical entities are related to multiple biomedical entities (i.e., N-to-M relations). When calculating the ranking of each symptom entity in the model output, we only consider the top 25 for convenience. If the symbol does not appear in the top 25, it will be calculated as 25, which can eliminate the deviation of the results caused by some abnormal points.

## 5.Results and Discussion

### 5.1 context VS no context

First, we report the changes of different models before and after adding EHR context. Following the work of BioLAMA, we also calculated acc@1 and acc@5 for three models, BERT, BioBERT and BioLM. We calculated three different sets of results, namely "no context", "segment1" and "avg all segments". Among them "no context" only inputs the prompt to the model; "segment1" is the prompt with a special context, because the segment1 context only contains a symptom of this target symptom, so there will be no noise symptoms; "avg all segments" is the average result of all kinds of context. As can be seen from Table 1, whether it is BERT, BioBERT or BioLM, the results of context ("segment1" and "avg all segments") are much better than the results of "no context". This result preliminarily proves that our method can greatly improve the low bound of the prompt paradigm in LAMA-like problems. As mentioned above, such improvement comes from the Transformer models paying more attention to the local context, and such local attention will make the model "soft constrict" its own output symptoms in some relatively reasonable candidates. This information all comes from the information retrieval work we did in the EHR notes, and the final improvement comes from the injection of this external knowledge. In addition, the results of "segment1" are also much better than the results of "avg all segments' '. Our explanation is that because the model does not have knowledge of all related symptoms for a certain disease, when other symptoms other than the target symptom appear, the model can output some of the symptoms it "remembers" according to its knowledge. But the other symptoms it "can't remember" are noises to the model, and these noises will bring down the ACC score. We will analyze the differences in the behavior of the model dealing with different symptoms through more detailed experiments in Section 5.2.

**Table 1.** Main experimental results on Context-Aware-BioLAMA

|  | no context |  | segment1 |  | avg all segments |  |
| --- | --- | --- | --- | --- | --- | --- |
|  | acc@1 | acc@5 | acc@1 | acc@5 | acc@1 | acc@5 |
| bert-base-cased | 0.013 | 0.161 | 0.108 | 0.226 | 0.069 | 0.198 |
| biobert-base-cased-v1.2 | 0.1 | 0.289 | 0.228 | 0.517 | 0.124 | 0.382 |
| Bio-LMs |  |  |  |  |  |  |
| roberta-base | 0.11 | 0.218 | 0.119 | 0.337 | 0.104 | 0.288 |
| RoBERTa-base-PM | 0.094 | 0.243 | 0.172 | 0.462 | 0.1 | 0.338 |
| RoBERTa-base-PM-Voc | 0.134 | 0.281 | 0.213 | 0.447 | 0.119 | 0.324 |
| RoBERTa-base-PM-M3 | 0.106 | 0.285 | 0.141 | 0.381 | 0.09 | 0.291 |
| RoBERTa-base-PM-M3-Voc | 0.102 | 0.273 | 0.142 | 0.336 | 0.074 | 0.233 |

Previous work has shown that the performance of different models on downstream tasks are closely related to the differences in training datasets and training vocabularies. We wanted to explore whether a similar relationship exists between the topics that the model contains knowledge. In this section, we used ACC@1 and ACC@5 results of 5 different bioLMs to answer this question. As shown in Table 1, most models' performance are in the order of segment1>avg all segments>no context, which is consistent with the results described in 5.1. Secondly, comparing the results of RoBERTa-base, RoBERTa-base-PM, and RoBERTa-base-PM-M3, we found that models that performed fine-tuning on domain specific datasets can get higher acc, indicating more relevant knowledge. Surprisingly, although both MIMIC III and UMass-EHRs datasets belong to EHR data, the results of RoBERTa-base-PM-M3 and RoBERTa-base-PM-M3-Voc are better than those of RoBERTa-base-PM and RoBERTa- The base-PM-Voc is poor. Through analysis, we found that because UMMS data belongs to general clinical EHR, and MIMIC III is more related to emergency room EHR, the entities (factual knowledge) contained in these two data are different. By observing the results of Voc and without Voc we find that a domain-specific vocabulary has a strong effect on such the LAMA problem.

**Table 2.** the changes of ACC@1 and ACC@5 before and after adding context. The three columns of each set of results represent: no context → segment 1, no context → avg all segments, segment 1 → avg all segments

|  | not acc@1 → acc@1 | | | acc@1 → not acc@1 | | | not acc@5 → acc@5 | | | acc@5 → not acc@5 | | |
| --- | --- | --- | --- | --- | --- | --- | --- | --- | --- | --- | --- | --- |
| bert-base-cased | 0.096 | 0.035 | 0.078 | 0.003 | 0.006 | 0.014 | 0.072 | 0.041 | 0.019 | 0.013 | 0.053 | 0.09 |
| **biobert-base-cased** | 0.163 | 0.05 | 0.159 | 0.038 | 0.067 | 0.018 | 0.259 | 0.118 | 0.023 | 0.038 | 0.102 | 0.228 |
| roberta-base | 0.055 | 0.034 | 0.056 | 0.048 | 0.043 | 0.04 | 0.157 | 0.095 | 0.046 | 0.042 | 0.047 | 0.114 |
| **RoBERTa-base-PM** | 0.122 | 0.042 | 0.114 | 0.047 | 0.062 | 0.018 | 0.266 | 0.12 | 0.037 | 0.054 | 0.082 | 0.212 |
| RoBERTa-base-PM-M3 | 0.092 | 0.04 | 0.083 | 0.059 | 0.071 | 0.019 | 0.18 | 0.094 | 0.042 | 0.09 | 0.123 | 0.161 |
| **RoBERTa-base-PM-Voc** | 0.142 | 0.041 | 0.141 | 0.065 | 0.079 | 0.026 | 0.223 | 0.084 | 0.039 | 0.063 | 0.088 | 0.202 |
| RoBERTa-base-PM-M3-Voc | 0.098 | 0.025 | 0.106 | 0.059 | 0.068 | 0.025 | 0.171 | 0.065 | 0.049 | 0.112 | 0.134 | 0.177 |

**5.2 Know it, or Say it?**

We further extend the experiments as shown in Table 1, we look at the ACC changes in three cases from "no context to segment 1", "no context to avg all segments", and "segment 1 to avg all segments". We looked at the ratio of four different changes. Taking "no context to segment 1" as an example, the first is the ratio of "no context" not in ACC@1 but "segment 1" entering ACC@1; the second is the ratio of "no context" in ACC@1 but "segment 1" is not in ACC@1; the third is the ratio of "no context" in ACC@5 but "segment 1" entering ACC@5; and the final one is the ratio of "no context" in ACC@5 but "segment 1" is not in ACC@5. We compute the same scores for "no context to avg all segments" and "segment 1 to avg all segments". As shown in Table 2, we found that BioBERT, RoBERTa-base-PM and RoBERTa-base-PM-Voc performed better in these experiments. For all models, when the model input goes from "no context" to "segment 1" or "no context" to "avg all segments", the ratio of "no acc" to "acc" is high, and much more than the ratio of "acc" to "not acc". For the case of "segment 1 to avg all segments", some models show that the ratio of "no acc" to "acc" is lower than the ratio of "acc" to "no acc". Consistent with the results of 5.1 and 5.2, the model can output more accurate symptoms for a disease through the added context, which is what we expect to see. But if we only give the model useful context (i.e. segment1), then a possible case is that the model just "copy it" simply because the symptoms appear in the context, rather than incorporating model's own knowledge (i.e. "know it") is selective to output what they think is true. So the results of "no context to avg all segments" can help us more accurately observe whether the model has some knowledge. Looking at Table 2 as a whole, the models have some knowledge about the relationship between diseases and symptoms, but not all, so they perform worse on avg all segments (containing noise symptoms) than segment1 (only with useful symptoms).

**Table 3.** For **Rank Change**, we calculate the average change in ranking of several groups of symptom entities each time a new segment (containing a new symptom and other contexts) is added. **The scores before and after the slash** come from whether the added segment contains a correct symptom or a wrong symptom. In the **Rank**, we calculated the average ranking of data with any length of context. We want to know, when a symptom is added to the input of the model along with the context, how the ranking of the newly added symptom and the existing symptoms

that are already in the input will be respectively. **Before and after slash** are the corresponding symptoms are correct or incorrect. **S: symptom, Sx: symptoms, Cor: correct, INCOR: incorrect**

|  | **Rank Change** (-ranking improve, +ranking decrease) | | | | **Rank**(smaller ranks higher) | |
| --- | --- | --- | --- | --- | --- | --- |
|  | target S | added S | COR Sx | INCOR Sx | added S | exist Sx |
| bert-base-cased | -0.986/0.783 | -6.567/-1.985 | -2.817/0.742 | 0.045/-0.589 | 18.42/23.01 | 18.89/23.27 |
| **biobert-base-cased** | -2.071/1.042 | -12.69/-6.053 | -5.625/1.061 | 0.231/-2.021 | 12.02/18.74 | 13.45/18.96 |
| roberta-base | -1.118/0.424 | -8.44/-3.253 | -4.083/0.415 | 0.029/-1.23 | 15.75/21.19 | 15.92/21.407 |
| **RoBERTa-base-PM** | -1.588/0.883 | -10.56/-3.762 | -5.206/0.911 | 0.094/-1.323 | 13.36/20.89 | 14.26/20.93 |
| RoBERTa-base-PM-M3 | -0.745/0.627 | -8.934/-3.065 | -4.055/0.65 | 0.057/-1.167 | 15.03/21.59 | 15.68/21.617 |
| RoBERTa-base-PM-M3-Voc | -0.52/0.656 | -7.035/-2.046 | -3.822/0.647 | 0.085/-0.901 | 16.56/22.44 | 17.37/22.483 |
| **RoBERTa-base-PM-Voc** | -1.313/0.835 | -9.399/-2.567 | -4.7/0.859 | 0.188/-0.984 | 14.18/22.01 | 15.33/22.13 |

We found that computing ACC@1 and ACC@5 alone does not provide a sufficient explanation of how the model handles context. We use some experiments about ranking to verify this. Specifically, when we expand the context window size, when the newly added segment contains the correct/wrong symptoms, how the ranking of different categories of symptom entities (target symptom, add symptom, correct symptoms, incorrect symptoms) changes in the output of the model. We calculated 6 scores below. When we add the EHR context to the prompt constructed for a certain triple (disease, has, target symptom), and we add the context segment by segment:

1. **Rank Change-target symptom**: changes in the ranking of target symptom when a new correct symptom is added (in the Table 3: before slash); changes in the ranking of target symptoms when a new incorrect symptom is added (after slash)
2. **Rank Change-added symptom**: changes in the ranking of newly added symptom when a correct symptom is added; changes in the ranking of newly added symptom when an incorrect symptom is added
3. **Rank Change-correct symptoms**: changes in the ranking of all existing correct symptoms in the input when a correct symptom is added; changes in the ranking of all existing correct symptoms in the input when an incorrect symptom is added
4. **Rank Change-incorrect symptoms**: changes in the ranking of all existing incorrect symptoms in the input when a correct symptom is added; changes in the ranking of all existing incorrect symptoms in the input when an incorrect symptom is added
5. **Rank-added symptom**: the rank of this new correct symptom when a new correct symptom is added; the rank of this new incorrect symptom when a new incorrect symptom is added
6. **Rank-existing symptoms**: the average rank of all existing correct symptoms in all data; the average rank of all existing incorrect symptoms in all data

All models have consistent trends in the **Rank change** and **Rank** of different categories of symptoms. That is, for the **Rank change** experiments, when a correct symptom is added in the new segment, models tend to increase the ranking of "target symptom", "added symptom" and "existing correct symptoms", and lower the ranking of "existing incorrect symptoms", which means that the model is able to group added correct symptom with other correct symptoms in context (including target symptom) based on its own knowledge, and then treat them in the same way, and existing incorrect symptoms in the opposite way; Correspondingly, when an incorrect symptom is added in the new segment, models tend to increase the ranking of "added symptom" and "existing incorrect symptoms", while lowering the ranking of "target symptom" and "existing correct symptoms". This also means that the model understands that newly added symptoms are incorrect and need to be dealt with together with existing incorrect symptoms, while remaining existing correct symptoms (including target symptoms) are dealt with in the opposite trend. Interestingly, we found that although the models tended to correctly classify the newly added symptom as existing correct symptoms or existing incorrect symptoms, the models always tended to improve the rank of the newly added symptom, regardless of whether it was correct or incorrect. We believe that the reason for this result is the local attention mechanism. The model naturally tends to mention the symptoms that have appeared in the context in the "filling in the blank" task, so the model tends to rank the added symptom higher. For the **Rank** experiment, observe the rank-added S column, when the newly added symptom is correct (slash left), the model will make the added symptom have a higher rank; observe the rank-exist S column, in all data, The model is also able to rank correct symptoms (slash left) higher. Based on the above explanations, when we compare the results of different models, we can find that the results in Table 3 are consistent with those in Table 2, with BioBERT, RoBERTa-base-PM and RoBERTa-base-PM-Voc leading all other models.

### 5.3 Cases Study and Analysis

We find some interesting cases to clarify the effect of dynamic-context-BioLAMA in more detail.

**Case1:** [segment 6], cough, [segment 4]shortness of breath [segment 2]or heart palpitations. No abdominal pain[segment 1], no nausea,[segment 3] vomiting, melena,[segment 5]hematochezia[segment 7], constipation or

    Prompt:Virus Diseases reflux disease has symptoms such as [Mask].

    Prompt → target rank: 25

    S1 + Prompt → target rank: 0  new symptom (correct) rank: 0

    S2 S1 + Prompt → target rank: 6  new symptom (correct) rank: 0

    S2 S1 S3 + Prompt → target rank: 7  new symptom (correct) rank: 4

    S4 S2 S1 S3 + Prompt → target rank: 5  new symptom (incorrect) rank: 25

    S4 S2 S1 S3 S5 + Prompt → target rank: 5  new symptom (incorrect) rank: 25

    S6 S4 S2 S1 S3 S5 + Prompt → target rank: 5  new symptom (correct) rank: 0

    S6 S4 S2 S1 S3 S5 S7 + Prompt → target rank: 5  new symptom (incorrect) rank: 25

**Case2:** [segment 2]HISTORY OF PRESENT ILLNESS: PATIENT comes in with his grandmother because of fever [segment 1]and belly pain. This fever started on the 12th. It is persisting, although he feels a little better today. He has had some abdominal pain, but no [segment 3]nausea or [segment 4]vomiting. He has [segment 5]generalized abdominal pain. He had a bit of a sore throat a couple of days ago. There has been no

    Prompt: Streptococcal Infections reflux disease has symptoms such as [Mask].

    Prompt → target rank: 1

    S1 + Prompt → target rank: 2  new symptom (correct) rank: 2

    S2 S1 + Prompt → target rank: 1  new symptom (correct) rank: 0

    S2 S1 S3 + Prompt → target rank: 1  new symptom (correct) rank: 4

    S2 S1 S3 S4 + Prompt → target rank: 1  new symptom (incorrect) rank: 3

    S2 S1 S3 S4 S5 + Prompt → target rank: 1  new symptom (incorrect) rank: 25

**Case3:** [segment 2]....... [segment 1]. He has very little energy. He is still able to shower and dress independently, but other than that he is mostly sedentary. He has felt this way for several weeks. Has not really improved since his discharge from the hospital. He is currently on antibiotics. He does not have any fevers or[segment 3] ……

    Prompt:Communicable Diseases reflux disease has symptoms such as [Mask].

    Prompt → target rank: 0

    S1 + Prompt → target rank: 5  new symptom (correct) rank: 5

In case 1 and case2, we can see that when the new symptom is correct, the model can often place them very early from the back; when the new symptom is incorrect, the model will place it in the back. This means that the model has the relevant knowledge. The results of target symptom are also very stable, adding correct symptom may lead to a drop in the ranking of target symptom (since the model may put correct symptom at the top, which is also reasonable), but adding incorrect symptoms will not influence target symptom's ranking since they will be put behind by the model. In case 2, the model placed the incorrect symptom third when adding segment 4, which is the most common source of error. Because the model does not have the knowledge about this disease and this added symptom, it mistakenly regards the added symptom as correct and places it very high.

Case 3 shows a different but interesting error. The EHR context mentions "not have any fevers". The model reads both "not" and "fever", so it is strongly influenced by the "not", and treat "fever" as an incorrect symptom. This is very interesting because local attention (context) and global attention (knowledge acquired by the model from pretrain) are the two main factors that affect the output of the model. The previous work hopes to use prompt to let the model output the knowledge it remembers according to global attention, but it is too difficult, so it can only be used as a low bound. Our work hopes to use local attention to "soft constrict" the scope of this knowledge to make it easier for the model to output the given knowledge. But this relationship between local attention and global attention needs more follow-up work to explore. For example, how to avoid the wrong influence of local context like case3.

**Conclusion**

In this work, we focus on the relationship between disease and symptoms, dynamically adding real EHR note data to prompts to observe how the knowledge output by the model changes. We show that adding EHR context can greatly improve the low bound of the prompt paradigm in the task of extracting model knowledge. Our experiments also show that the knowledge possessed by the model can distinguish the correct knowledge (correct symptoms) from the noise knowledge (incorrect symptoms) in the EHR, and such distinguishing ability can also be used as a metric of the amount of knowledge possessed by the model.

**Acknowledgments**